\documentclass[a4paper,onecolumn,11pt]{quantumarticle}
\pdfoutput=1
\usepackage[utf8]{inputenc}
\usepackage[T1]{fontenc}
\usepackage[english]{babel}
\usepackage{amsmath,amssymb,amsthm,mathtools}
\usepackage{braket,dsfont,mathrsfs}
\usepackage{enumitem}
\usepackage{hyperref}
\usepackage{microtype}
\hypersetup{colorlinks=true,citecolor=blue,linkcolor=blue,urlcolor=blue}
\usepackage[capitalize,nameinlink]{cleveref}
\numberwithin{equation}{section}

\theoremstyle{plain}
\newtheorem{theorem}{Theorem}[section]
\newtheorem{lemma}[theorem]{Lemma}
\newtheorem{proposition}[theorem]{Proposition}
\newtheorem{corollary}[theorem]{Corollary}
\theoremstyle{definition}
\newtheorem{definition}[theorem]{Definition}
\newtheorem{example}[theorem]{Example}
\theoremstyle{remark}
\newtheorem{remark}[theorem]{Remark}

\newcommand{\1}{\mathds{1}}

\newcommand{\cX}{\mathcal{X}}
\newcommand{\cY}{\mathcal{Y}}
\newcommand{\hi}{\mathcal{H}}
\newcommand{\state}{\mathcal{S}}
\newcommand{\PP}{\mathsf{P}}
\newcommand{\Q}{\mathsf{Q}}
\newcommand{\R}{\mathsf{R}}
\newcommand{\M}{\mathcal{M}}
\newcommand{\Rng}{\mathscr{R}}
\newcommand{\Eff}{\mathscr{E}}
\newcommand{\B}{\mathcal{B}}

\newcommand{\lmin}{\lambda_{\min}}
\newcommand{\lmax}{\lambda_{\max}}
\newcommand{\Tr}[1]{\operatorname{Tr}\!\left[#1\right]}
\newcommand{\ketbra}[2]{\ket{#1}\!\bra{#2}}

\renewcommand{\le}{\leqslant}
\renewcommand{\ge}{\geqslant}
\renewcommand{\leq}{\leqslant}
\renewcommand{\geq}{\geqslant}

\title{From outcome exclusion to measurement simulation}
\author{Kensei Torii}
\email{toriikensei@nagoya-u.jp}
\author{Jiaxi Kuang}
\author{Francesco Buscemi}
\email{buscemi@nagoya-u.jp}
\affiliation{Department of Mathematical Informatics, Graduate School of Informatics, Nagoya University}

\begin{document}
\maketitle

\begin{abstract}
While quantum state exclusion has received considerable attention, the analogous task of measurement outcome exclusion remains less explored. Here we develop a quantitative description of outcome exclusion and extend the resulting framework to obtain necessary and sufficient criteria for measurement simulation by quantum pre-processing and classical post-processing. Our analysis combines two approaches. Algebraic exclusion indices quantify normalized minimum errors in communication tasks with forbidden outputs and yield lower bounds on the number of classical messages required to reproduce the task statistics. Random exclusion profiles provide a statistical description in terms of optimized hypothesis-testing errors and are monotone under both types of processing. This connects outcome exclusion with statistical comparison in the spirit of Blackwell, linking the ability to suppress measurement outcomes to the possibility of simulating one measurement from another.
\end{abstract}

\section{Introduction}

A measurement is usually judged by what its outcomes allow us to infer about the system being measured. Two limiting possibilities are particularly transparent: an outcome may occur with certainty, or it may be ruled out with certainty. For a finite-outcome quantum measurement described by a POVM $\PP=\{P_x\}_{x\in\cX}$, the former means that some input state produces $x$ with probability one, whereas the latter means that some input state makes $x$ impossible. Here, we regard the ability to rule out one or more outcomes with certainty as a desirable property, or ``resource'', of a measurement.

There is, however, a difficulty with a formulation based only on outcome labels. Classical post-processing may split one outcome into several labels, merge different outcomes, or introduce labels that never occur. In particular, an outcome corresponding to a null effect is impossible for every preparation, so excluding it says nothing about the measurement. Hence, the question is how to compare exclusion capabilities in a way that has a direct operational meaning and remains well behaved under processing of the classical output.

We approach this question in two steps: an algebraic one, related to communication cost, and a statistical one, related to hypothesis testing. In the algebraic step, for every set $\B\subseteq\cX$ of forbidden outcomes, we consider the smallest probability of obtaining an outcome in $\B$,
\begin{equation}
    \epsilon_{\B}(\PP)=\lambda_{\min}\!\left(\sum_{x\in\B}P_x\right).
\end{equation}
Normalized sums over all sets of size $r$ define the algebraic exclusion indices $\epsilon_r(\PP)$. The condition $\epsilon_r(\PP)=0$ holds exactly when every set of $r$ outcomes can be excluded with certainty. As $r$ increases from $1$ to $|\cX|-1$, this requirement ranges from excluding any individual outcome to producing any specified outcome with certainty, namely sharpness~\cite{Buscemi2024-sharpness}. These indices give the normalized minimum error in a communication task whose goal is to produce an outcome outside a prescribed forbidden set; comparison with classical protocols then yields lower bounds on the number of messages needed to reproduce the task statistics~\cite{Frenkel_Weiner_2015}. However, their dependence on the chosen outcome alphabet prevents monotonicity under arbitrary classical post-processing.

To overcome this limitation, we take a statistical approach, replacing fixed forbidden sets by randomized binary events. Such an event is specified by a function $f:\cX\to[0,1]$: after outcome $x$, the event is declared to occur with probability $f_x$. Its probability on input $\rho$ is $\Tr{\rho P_f}$, where $P_f=\sum_xf_xP_x$. To avoid a trivial minimization over events that never occur, we require that the event have probability at least $w>0$ for some preparation. Minimizing its probability over another preparation and then over the event gives
\begin{equation}
    E^w(\PP)
    :=\min_{f:\,\lambda_{\max}(P_f)\geq w}\lambda_{\min}(P_f),
    \qquad 0<w\leq1.
\end{equation}
We call the family $\{E^w(\PP)\}_{w\in(0,1]}$ the \textit{random exclusion profile} of the POVM $\PP$. It is exactly an optimized hypothesis-testing quantity for the quantum-to-classical channel associated with $\PP$, and it is monotone under quantum pre-processing and arbitrary classical post-processing.

The statistical formulation also connects exclusion with simulation. We prove that comparison of the random exclusion profiles is equivalent to comparison of all binary post-processings through inclusion of their statistical ranges, namely the sets of output distributions attainable by varying the input state. Replacing binary post-processings by $k$-outcome ones gives a hierarchy of preorders, with the number of retained outcomes specifying how much of the measurement statistics is compared. Range inclusion first yields positive unital maps on the operator systems spanned by the post-processed effects. When both measurements are supplemented with the same informationally complete ancillary measurement, the corresponding comparison yields \textit{completely} positive maps. Retaining all outcomes of the target measurement then gives a necessary and sufficient criterion for simulation by a single quantum pre-processing channel and a classical post-processing matrix. The positive and completely positive simulation hierarchies coincide at the binary level; in particular, comparison of the random exclusion profiles already gives a complete simulation criterion when the target measurement has two outcomes.

The range criteria underlying this construction originate in the theory of clean POVMs \cite{buscemi-2005-clean-POVMs}, which also settles the case of two outcomes. Classical post-processing defines the nonideality preorder of Ref.~\cite{Martens_deMuynck_1990} and underlies the resource theories of measurement informativeness \cite{Skrzypczyk_Linden_2019,Guff_2021}; in that setting, the success probabilities in all state discrimination games \cite{Buscemi_2016,Buscemi_2018} and the error probabilities in all state exclusion games \cite{Ducuara_Skrzypczyk_2019} form complete families of monotones. Ref.~\cite{Buscemi2024-sharpness} instead compares measurements through quantum pre-processing and mixing with trivial POVMs.

The paper is organized as follows. Section~\ref{sec:preliminaries} fixes notation and recalls the relevant spectral facts. Section~\ref{sec:An algebraic hierarchy of outcome exclusion} introduces the algebraic profile and its communication interpretation. Section~\ref{sec:A statistical hierarchy of outcome exclusion} develops random exclusion and its binary range characterization. Section~\ref{sec:Simulation hierarchy of POVMs} constructs the simulation hierarchies and characterizes simulation. Section~\ref{sec:Discussion} discusses open questions.

\section{Preliminaries}\label{sec:preliminaries}

All Hilbert spaces and outcome sets are finite. We write $\mathcal L(\hi)$ for the linear operators on $\hi$, $\state(\hi)$ for its density operators, and $\1$ for the identity operator, with its space understood from context. A POVM on $\hi$ with outcome set $\cX$ is a family $\PP=\{P_x\}_{x\in\cX}$ such that $P_x\geq0$ and $\sum_xP_x=\1$. Its associated quantum-to-classical channel is
\begin{equation}
    \M_{\PP}(\rho)=\sum_{x\in\cX}\Tr{\rho P_x}\ketbra{x}{x},
\end{equation}
where $\{\ket{x}\}_{x\in\cX}$ is an orthonormal basis for the classical output register. The statistical range of $\PP$ is the compact convex set
\begin{equation}\label{eq:povm-range}
    \Rng(\PP):=\left\{\bigl(\Tr{\rho P_x}\bigr)_{x\in\cX}:\rho\in\state(\hi)\right\}
    \subseteq\Delta(\cX),
\end{equation}
where $\Delta(\cX)$ is the probability simplex on $\cX$.

\begin{definition}[Trivial and sharp POVMs; excludable outcomes]\label{def:basic-povms}
Let $\PP=\{P_x\}_{x\in\cX}$ be a POVM.
\begin{enumerate}
    \item $\PP$ is trivial if $P_x=p_x\1$ for a probability distribution $p$ on $\cX$.
    \item $\PP$ is sharp, in the operational sense of Ref.~\cite{Buscemi2024-sharpness}, if $\lambda_{\max}(P_x)=1$ for every $x\in\cX$.
    \item An outcome $x$ is excludable if $P_x$ has a non-trivial kernel.
\end{enumerate}
\end{definition}

The definition of sharpness is operational and does not require the effects to be projections. If $P_x$ has eigenvalue one, then $\sum_{y\ne x}P_y=\1-P_x$ has a non-trivial kernel. Thus the input that produces $x$ with certainty excludes every other outcome. In the case of at least two outcomes, every nonzero effect of a sharp POVM therefore has a non-trivial kernel. A null effect is trivially excludable; its role will be removed by the probability threshold constraint in Section~\ref{sec:A statistical hierarchy of outcome exclusion}.

For every self-adjoint operator $A$,
\begin{equation}
    \min_{\rho\in\state(\hi)}\Tr{\rho A}=\lambda_{\min}(A),
    \qquad
    \max_{\rho\in\state(\hi)}\Tr{\rho A}=\lambda_{\max}(A).
\end{equation}
Moreover, if $\Phi$ is positive and unital on an operator system containing $A$, then
\begin{equation}\label{eq:cpu-spectrum-shrink}
    \lambda_{\min}(\Phi(A))\geq\lambda_{\min}(A),
    \qquad
    \lambda_{\max}(\Phi(A))\leq\lambda_{\max}(A).
\end{equation}
Indeed, apply $\Phi$ to $\lambda_{\min}(A)\1\leq A\leq\lambda_{\max}(A)\1$. Here an operator system is a self-adjoint complex linear subspace containing the identity, equipped with the positivity inherited from its ambient matrix algebra~\cite{Paulsen_2002}. In particular, the complex linear span of the effects of a POVM is an operator system.

Quantum pre-processing by a completely positive trace-preserving map $\Lambda:\mathcal L(\hi_{\Q})\to\mathcal L(\hi_{\PP})$, where $\hi_{\Q}\,(\hi_{\PP})$ is the Hilbert space associated with $\Q\,(\PP)$, acts on the effects of $\PP$ through the completely positive unital map $\Phi=\Lambda^{\dagger}$. Classical post-processing acts through a stochastic matrix $\mu(y|x)$, with $\mu(y|x)\geq0$ and $\sum_y\mu(y|x)=1$, and produces effects $Q_y=\sum_x\mu(y|x)P_x$. We keep the domains of positive maps explicit, since positivity on an operator system alone need not imply the existence of a positive extension to the full matrix algebra~\cite{Arveson_1969,Paulsen_2002}.

\section{Exclusion of $r$ outcomes: an algebraic hierarchy}
\label{sec:An algebraic hierarchy of outcome exclusion}

Predicting one outcome with certainty requires excluding all the others. This observation suggests that single-outcome exclusion and sharpness should be regarded as the endpoints of a hierarchy, whose intermediate levels concern the simultaneous exclusion of several outcomes. This parallels, on the side of measurements, the $m$-state conclusive exclusion of sets of states~\cite{Bandyopadhyay_2014,Stratton_2024}, which ranges from the exclusion of a single state to perfect discrimination. We first construct this hierarchy from the spectra of sums of POVM effects, and then identify its indices with normalized errors in a communication task with forbidden outputs.

\subsection{Algebraic exclusion profile}

Let $\PP=\{P_x\}_{x\in\cX}$ be a POVM and write $n:=|\cX|\ge2$. For a subset $\B\subseteq\cX$, define
\begin{equation}
    P_\B:=\sum_{x\in\B}P_x,
    \qquad
    \epsilon_\B(\PP):=\lmin(P_\B).
    \label{eq:subset-exclusion-index}
\end{equation}
The quantity $\epsilon_\B(\PP)$ is the minimum probability of obtaining an outcome in $\B$, optimized over the input state. Hence it vanishes precisely when a suitable preparation excludes all outcomes in $\B$ simultaneously. To compare sets of the same cardinality without privileging particular labels, we sum these quantities over all subsets of size $r$ and normalize by the value attained by a trivial POVM.

\begin{definition}[Algebraic exclusion profile]
\label{def:epsilon_r}
For $r=1,\ldots,n-1$, the algebraic exclusion index of $\PP$ is
\begin{equation}
    \epsilon_r(\PP)
    :=\frac{1}{c_r}\sum_{\B\subseteq\cX:\,|\B|=r}\epsilon_\B(\PP),
    \qquad
    c_r:=\binom{n-1}{r-1}.
    \label{eq:def-epsilon-r}
\end{equation}
The indexed family $(\epsilon_r(\PP))_{r=1}^{n-1}$ is called the algebraic exclusion profile of $\PP$.
\end{definition}

Smaller values correspond to a greater ability to exclude outcomes; the operational meaning of the normalization will be established in \cref{subsec:forbidden-output-task}.

\begin{proposition}[Basic properties of the algebraic profile]
\label{prop:epsilon-basic}
Let $\PP=\{P_x\}_{x\in\cX}$ be a POVM with $n\ge2$ outcomes. The following statements hold:
\begin{enumerate}
    \item $0\le\epsilon_r(\PP)\le1$ for every $r=1,\ldots,n-1$.
    \item $\epsilon_r(\PP)=0$ if and only if every subset of $r$ outcomes can be excluded with certainty.
    \item $\epsilon_{n-1}(\PP)=0$ if and only if $\PP$ is sharp.
    \item For any $r=1,\ldots,n-1$, $\epsilon_r(\PP)=1$ if and only if $\PP$ is trivial.
    \item If $1\le r<r'\le n-1$, then $\epsilon_r(\PP)\le\epsilon_{r'}(\PP)$.
\end{enumerate}
\end{proposition}

\begin{proof}
All summands in \cref{eq:def-epsilon-r} are nonnegative. For the upper bound, notice that each outcome belongs to exactly $c_r$ subsets of size $r$. Thus, for every unit vector $\psi$,
\begin{equation}
    \sum_{\B\subseteq\cX:|\B|=r}\lmin(P_\B)
    \le\sum_{\B\subseteq\cX:|\B|=r}\bra{\psi}P_\B\ket{\psi}
    =c_r\bra{\psi}\1\ket{\psi}
    =c_r.
    \label{eq:epsilon_upper_bound}
\end{equation}
This proves the first statement.

The sum defining $\epsilon_r$ vanishes if and only if $\lmin(P_\B)=0$ for every $r$-element subset $\B$. Since all effects are positive, a unit vector $\psi$ belongs to $\ker P_\B$ if and only if it belongs to $\ker P_x$ for every $x\in\B$. This is precisely the condition for excluding the whole set $\B$ and proves the second statement.

For $r=n-1$, the relevant subsets are the complements of single outcomes. Therefore
\begin{equation}
    \epsilon_{n-1}(\PP)
    =\frac{1}{n-1}\sum_{x\in\cX}\lmin(\1-P_x)
    =\frac{1}{n-1}\sum_{x\in\cX}\bigl(1-\lmax(P_x)\bigr).
    \label{eq:sharpness-exclusion-endpoint}
\end{equation}
Every summand is nonnegative, so the expression vanishes exactly when every effect has eigenvalue one, namely, when $\PP$ is sharp.

If $\PP$ is trivial, with $P_x=p_x\1$, then
\[
    \epsilon_r(\PP)
    =\frac{1}{c_r}\sum_{\B\subseteq\cX:|\B|=r}\sum_{x\in\B}p_x
    =1.
\]
Conversely, suppose $\epsilon_r(\PP)=1$. The inequality in \cref{eq:epsilon_upper_bound} then saturates for every unit vector $\psi$. Since each difference
$\bra{\psi}P_\B\ket{\psi}-\lmin(P_\B)$ is nonnegative, each difference must vanish. Hence every sum $P_\B$ with $|\B|=r$ is proportional to the identity. For $r=1$, this already proves triviality. For $r>1$, choose distinct $x,y\in\cX$ and a set $\mathcal C\subseteq\cX\setminus\{x,y\}$ of size $r-1$, which is possible because $r\le n-1$. Both $P_x+P_{\mathcal C}$ and $P_y+P_{\mathcal C}$ are proportional to the identity, and thus so is $P_x-P_y$. Since this holds for all $x,y$ and $\sum_xP_x=\1$, every $P_x$ is proportional to the identity as well.

Finally, it suffices to prove monotonicity between consecutive levels. For a set $\mathcal C\subseteq\cX$ of size $r+1$,
\[
    \sum_{\B\subset\mathcal C:\,|\B|=r}P_\B=rP_{\mathcal C}.
\]
Superadditivity of the minimum eigenvalue gives
\[
    \sum_{\B\subset\mathcal C:\,|\B|=r}\lmin(P_\B)
    \le r\lmin(P_{\mathcal C}).
\]
Upon summing over all such $\mathcal C$, each $r$-element subset appears exactly $n-r$ times, so that
\[
    (n-r)\sum_{\B\subseteq\cX:|\B|=r}\lmin(P_\B)
    \le r\sum_{\mathcal C\subseteq\cX:|\mathcal C|=r+1}\lmin(P_{\mathcal C}).
\]
Using $c_{r+1}=(n-r)c_r/r$, this becomes $\epsilon_r(\PP)\le\epsilon_{r+1}(\PP)$.
\end{proof}

In terms of the measurement range $\Rng(\PP)$, the condition $\epsilon_r(\PP)=0$ means that this range meets every face of the probability simplex specified by setting $r$ coordinates to zero. In particular, sharpness requires the range to contain every vertex of the simplex (and thus the whole simplex).

\begin{remark}[Endpoints of the algebraic profile]
\label{rem:epsilon-endpoints}
The two extreme indices coincide, up to an affine reparametrization, with known quantifiers of measurement informativeness. Since $c_1=1$,
\begin{equation}
    \epsilon_1(\PP)=\sum_{x\in\cX}\lmin(P_x)=1-\mathrm{WoI}(\PP),
    \label{eq:epsilon1-woi}
\end{equation}
where the weight of informativeness $\mathrm{WoI}(\PP)$ is the smallest $w\ge0$ such that $P_x=wN_x+(1-w)q_x\1$ for some POVM $\{N_x\}_{x\in\cX}$ and some probability distribution $q$ on $\cX$~\cite{Ducuara_Skrzypczyk_2019,Ducuara_Skrzypczyk_2020}. Indeed, such a decomposition exists if and only if $(1-w)q_x\le\lmin(P_x)$ for every $x$, and summing over $x$ shows that the largest admissible value of $1-w$ is $\sum_x\lmin(P_x)$. At the other endpoint, \cref{eq:sharpness-exclusion-endpoint} reads
\begin{equation}
    \epsilon_{n-1}(\PP)=1-\frac{\mathrm{RoM}(\PP)}{n-1},
    \qquad
    \mathrm{RoM}(\PP):=\sum_{x\in\cX}\lmax(P_x)-1,
    \label{eq:epsilon-n-1-rom}
\end{equation}
where $\mathrm{RoM}(\PP)$ is the robustness of measurement~\cite{Skrzypczyk_Linden_2019}. Both quantities have a known operational meaning: $\epsilon_1(\PP)$ is the smallest ratio, over all ensembles of states, between the error probability attained with $\PP$ in a state exclusion game and the one attained without measuring~\cite{Ducuara_Skrzypczyk_2019,Uola_2020}, while $1+\mathrm{RoM}(\PP)$ is the largest such ratio between success probabilities in state discrimination games~\cite{Skrzypczyk_Linden_2019}. The intermediate indices $\epsilon_r$, with $1<r<n-1$, interpolate between these two quantifiers.
\end{remark}

\subsection{Monotonicity of the algebraic hierarchy}

Quantum pre-processing restricts the states available at the input of a measurement, and hence cannot improve exclusion. In the Heisenberg picture, it acts on effects by a completely positive unital map. The same spectral argument in fact applies to any positive unital map, including maps defined only on an operator system containing the effects.

For classical post-processing, a natural class preserving the uniform treatment of outcome labels is given by bistochastic matrices on $\cX$. Such a matrix satisfies
\[
    \sum_y\mu(y|x)=1,
    \qquad
    \sum_x\mu(y|x)=1,
\]
and transforms the measurement according to
\begin{equation}
    Q_y=\sum_x\mu(y|x)P_x.
    \label{eq:def_Q}
\end{equation}
The exclusion indices do not decrease under either operation.

\begin{proposition}[Monotonicity of the algebraic profile]
\label{prop:epsilon-monotone}
Let $\PP=\{P_x\}_{x\in\cX}$ have $n\ge2$ outcomes.
\begin{enumerate}
    \item If $\Phi$ is positive and unital and $\PP'=\{\Phi(P_x)\}_{x\in\cX}$, then
    \[
        \epsilon_r(\PP')\ge\epsilon_r(\PP)
        \qquad(r=1,\ldots,n-1).
    \]
    \item If $\Q$ is obtained from $\PP$ by a bistochastic post-processing as in \cref{eq:def_Q}, then
    \[
        \epsilon_r(\Q)\ge\epsilon_r(\PP)
        \qquad(r=1,\ldots,n-1).
    \]
\end{enumerate}
\end{proposition}

\begin{proof}
For the first statement, positivity and unitality imply
\[
    \lmin\!\left(\Phi(P_\B)\right)\ge\lmin(P_\B)
\]
for every $\B$, by \cref{eq:cpu-spectrum-shrink}. Summation and normalization give the result.

For the second statement, the Birkhoff--von Neumann theorem (see e.g. \cite{Bhatia1997}) gives a decomposition
\begin{equation}
    \mu(y|x)=\sum_\pi\alpha_\pi\delta_{y,\pi(x)},
    \qquad
    \alpha_\pi\ge0,
    \qquad
    \sum_\pi\alpha_\pi=1,
    \label{eq:mu_permutarion}
\end{equation}
where $\pi$ ranges over permutations of $\cX$. Consequently,
\[
    Q_\B=\sum_\pi\alpha_\pi P_{\pi^{-1}(\B)},
    \qquad
    \lmin(Q_\B)\ge\sum_\pi\alpha_\pi\lmin(P_{\pi^{-1}(\B)}).
\]
Since $\B\mapsto\pi^{-1}(\B)$ permutes the subsets of size $r$, summing the last inequality gives
\[
    \epsilon_r(\Q)
    \ge\sum_\pi\alpha_\pi\epsilon_r(\PP)
    =\epsilon_r(\PP).
\]
\end{proof}

The restriction on classical post-processing is relevant even when the number of labels is kept fixed. For example, let $A$ and $B$ be complementary rank-one projections on a qubit, and consider
\[
    \PP=(A/2,A/2,B/2,B/2),
    \qquad
    \Q=(A,0,B/2,B/2).
\]
The latter is obtained by merging the first two outcomes and retaining the second label as a null outcome. Nevertheless, direct evaluation gives $\epsilon_2(\PP)=2/3$ and $\epsilon_2(\Q)=1/3$. Thus unrestricted post-processing can improve an algebraic exclusion index solely by changing the representation of the classical output. The statistical formulation of \cref{sec:A statistical hierarchy of outcome exclusion} will restore monotonicity under arbitrary classical post-processing.

\subsection{Operational meaning in a communication task with forbidden outputs}
\label{subsec:forbidden-output-task}

The vanishing of $\epsilon_r$ already has a direct interpretation. The question then arises as to what its positive values quantify. To answer this question, consider a referee who chooses an $r$-element subset $\B\subseteq\cX$ uniformly at random and reveals it only to Alice. This subset specifies the forbidden outputs. Alice prepares a state $\rho_\B$ and sends it to Bob, who performs the fixed measurement $\PP$ and reports its outcome $x$. They lose if $x\in\B$.  This is a prepare-and-measure task of the kind introduced in Refs.~\cite{Heinosaari_Kerppo_2019,Heinosaari_2020} under the name of \textit{communication of partial ignorance}, in which the sender is told which alternatives the receiver should avoid; the difference is that here Bob's measurement is fixed, and only Alice's preparations are optimized. Related exclusion games are known to separate quantum from classical communication~\cite{Perry_2015,Havlicek_Barrett_2020}.

For a family of preparations $(\rho_\B)_{\B:|\B|=r}$, the average losing probability is
\begin{equation}
    L_r\bigl(\PP,(\rho_\B)_\B\bigr)
    :=\binom nr^{-1}\sum_{\B\subseteq\cX:|\B|=r}\sum_{x\in\B}\Tr{\rho_\B P_x},
    \label{eq:forbidden-output-loss}
\end{equation}
and its optimum is
\begin{equation}
    L_r(\PP):=\min_{(\rho_\B)_\B}L_r\bigl(\PP,(\rho_\B)_\B\bigr).
    \label{eq:optimal-forbidden-output-loss}
\end{equation}
Alice may choose the preparation independently for each forbidden set. Hence
\begin{equation}
\begin{split}
    L_r(\PP)
    &=\binom nr^{-1}\sum_{\B\subseteq\cX:|\B|=r}
        \min_{\rho_\B\in\state(\hi)}\Tr{\rho_\B P_\B}\\
    &=\binom nr^{-1}\sum_{\B\subseteq\cX:|\B|=r}\lmin(P_\B)
     =\frac{r}{n}\epsilon_r(\PP).
\end{split}
\label{eq:forbidden-output-optimum}
\end{equation}
Each minimum is attained by a state supported on the eigenspace of the smallest eigenvalue of $P_\B$.

In the absence of communication, Bob's output is independent of $\B$. Every fixed output belongs to exactly $\binom{n-1}{r-1}$ forbidden sets, so every such strategy has the same losing probability,
\begin{equation}
    L_r^{\mathrm{no\,comm}}=
    \frac{\binom{n-1}{r-1}}{\binom nr}
    =\frac rn.
    \label{eq:no-communication-loss}
\end{equation}
Combining \cref{eq:forbidden-output-optimum,eq:no-communication-loss}, we obtain
\begin{equation}
    \epsilon_r(\PP)=\frac{L_r(\PP)}{L_r^{\mathrm{no\,comm}}}.
    \label{eq:operational-meaning-epsilon-r}
\end{equation}
In other words, $\epsilon_r$ is the minimum losing probability with the measurement $\PP$, expressed relative to the losing probability without communication. This interpretation also explains why trivial measurements have index one.

We next compare this task with classical communication. Suppose Alice may send one among $k$ classical messages, and that Alice and Bob have access to shared randomness independent of the forbidden set. A randomized protocol is an average of deterministic protocols, so it cannot lower the minimum of this linear loss function. For a deterministic protocol, Bob associates an output $x_j$ with each message $j$. Alice can avoid a loss exactly when at least one of these outputs lies outside $\B$. It is therefore optimal to choose $k$ distinct outputs, whenever $k\le n$, and the remaining loss is the probability that all of them belong to the forbidden set. Thus
\begin{equation}
    L_{n,r,k}^{\mathrm{cl}}
    =\frac{\binom{n-k}{r-k}}{\binom nr}
    =\frac{\binom rk}{\binom nk}
    \qquad(1\le k\le r),
    \label{eq:classical-forbidden-output-benchmark}
\end{equation}
whereas $L_{n,r,k}^{\mathrm{cl}}=0$ for $r<k\le n$. Equivalently, the last ratio applies throughout $1\le k\le n$ with the convention $\binom rk=0$ for $k>r$.

Fix $r$ and choose an optimal preparation $\rho_\B$ for each forbidden set $\B$. The resulting conditional probabilities $p(x|\B)=\Tr{\rho_\B P_x}$ define a finite probability table whose average losing probability is $L_r(\PP)=(r/n)\epsilon_r(\PP)$. Any classical protocol reproducing this table exactly must also reproduce its average losing probability. Hence, if the protocol uses one among $k$ classical messages, with shared randomness allowed, its error cannot be smaller than the classical optimum in \cref{eq:classical-forbidden-output-benchmark}. It follows that
\begin{equation}
    \epsilon_r(\PP)
    \ge\frac nr\frac{\binom rk}{\binom nk}
    \qquad(k\le r\le n-1).
    \label{eq:epsilon-classical-message-bound}
\end{equation}
A violation therefore certifies that more than $k$ classical messages are necessary to reproduce this finite probability table. In particular, it rules out any classical protocol reproducing the statistics of $\PP$ for every preparation with the same message budget: in the terminology of Refs.~\cite{DallArno_2017,Doolittle_Chitambar_2021}, it shows that the signaling dimension of $\PP$, namely the minimum number of classical messages that, assisted by shared randomness, reproduce all its input--output statistics, exceeds $k$. The converse does not follow: satisfying \cref{eq:epsilon-classical-message-bound} only constrains the average loss and does not by itself guarantee that all conditional probabilities can be reproduced. 

On the other hand, the dimension of the Hilbert space gives an upper bound on the required number of classical messages. Indeed, the theorem of Frenkel and Weiner~\cite{Frenkel_Weiner_2015} guarantees that, for any finite family of preparations and any fixed POVM on a $d$-dimensional Hilbert space, the entire probability table can be reproduced with $d$ classical messages and shared randomness. Thus the simulation assumed above always exists for $k=d$, and every such POVM satisfies
\[
    \epsilon_r(\PP)
    \ge\frac nr\frac{\binom rd}{\binom nd}
    =\frac{\binom{n-d}{r-d}}{\binom{n-1}{r-1}},
    \qquad d\le r\le n-1.
\]
This is a universal restriction on $d$-dimensional measurements and coincides with the normalized form of the subset inequality derived in Section~3 of Ref.~\cite{Frenkel_Weiner_2015}. In other words, the possibility of simulating the statistics with $d$ classical messages limits how small the exclusion error of a $d$-dimensional measurement can be.

Notice that ruling out a $k$-message simulation does not require perfect exclusion: it is enough for $\epsilon_r(\PP)$ to lie below the positive threshold in \cref{eq:epsilon-classical-message-bound}. The following example shows that this can occur even when all effects are full rank, and that the resulting communication bound can be stronger than the one obtained from ordinary guessing.

\begin{example}[A communication witness beyond ordinary guessing]
\label{ex:noisy-uniform-pairs}
Let $\cX=\{1,2,3,4\}$ and let $\hi=\mathbb C^6$ have orthonormal basis $(\ket{S})_{S\subseteq\cX:\,|S|=2}$. Define
\begin{equation}
    P_x^{(0)}:=\frac12\sum_{S\subseteq\cX:\,|S|=2,\ x\in S}\ketbra{S}{S},
    \label{eq:uniform-pair-povm}
\end{equation}
and introduce uniform noise by setting
\begin{equation}
    P_x^{(t)}:=(1-t)P_x^{(0)}+\frac t4\1,
    \qquad 0\le t\le1.
    \label{eq:noisy-uniform-pair-povm}
\end{equation}
For a forbidden pair $\B$, the preparation $\ketbra{\cX\setminus\B}{\cX\setminus\B}$ attains the minimum eigenvalue of $P_\B^{(t)}$, namely $t/2$. Consequently,
\[
    L_2(\PP^{(t)})=\frac t2,
    \qquad
    \epsilon_2(\PP^{(t)})=t.
\]
Since $L_{4,2,2}^{\mathrm{cl}}=1/6$, the interval $0<t<1/3$ gives a violation of the bound for two classical messages. Thus at least three messages are needed to reproduce these statistics, although all effects are positive definite and no nonempty set of outcomes can be excluded with certainty.

For comparison, consider the ordinary task in which Alice is given a uniformly distributed label $x\in\cX$ and Bob succeeds by reporting that same label. With the fixed measurement $\PP$, its optimal success probability is $\frac14\sum_x\lmax(P_x)$. A protocol using $k$ classical messages has success probability at most $k/4$, even with shared randomness, because a deterministic decoding can correctly identify at most $k$ of the four labels. Hence the associated guessing bound for simulation is $\sum_x\lmax(P_x)\le k$. In the present example,
\[
    \sum_{x=1}^{4}\lmax(P_x^{(t)})=2-t\le2.
\]
The guessing bound therefore does not rule out two messages, whereas the intermediate exclusion index does.
\end{example}

\section{Exclusion of randomized events: a statistical hierarchy}
\label{sec:A statistical hierarchy of outcome exclusion}

The preceding hierarchy treats all subsets of a given cardinality on the same footing. However, as we already noticed, cardinality refers to the chosen outcome alphabet, and arbitrary classical post-processing need not preserve that structure. To formulate an exclusion comparison that is stable under such post-processing, we now regard an excluded event as a binary statistical test on the measurement output. Deterministic subsets remain special cases, but randomized tests allow us to compare events through their probabilities rather than through the number of labels used to represent them.

\subsection{Random exclusion games}

A classical effect is a function $f:\cX\to[0,1]$ representing a randomized event. After observing the outcome $x$, a referee declares a designated event to have occurred with probability $f_x$. The corresponding quantum effect is
\begin{equation}
    P_f:=\sum_x f_xP_x,
    \label{eq:def_P_f}
\end{equation}
and the set of all effects obtainable in this way is
\begin{equation}
    \Eff(\PP):=\left\{P_f:f:\cX\to[0,1]\right\}.
    \label{eq:effect-range}
\end{equation}
We call this set the effect range of $\PP$; operationally, it contains all binary tests available after the measurement.

\begin{definition}[Random exclusion game]
\label{def:random-exclusion-game}
For a classical effect $f:\cX\to[0,1]$, the player chooses the input state and loses when the event specified by $f$ occurs. The optimal losing probability is
\begin{equation}
    E_f(\PP):=\min_{\rho\in\state(\hi)}\Tr{\rho P_f}
    =\lmin(P_f).
    \label{eq:fixed-random-exclusion-index}
\end{equation}
\end{definition}

For the indicator function of a subset $\B$, this reduces to $E_f(\PP)=\epsilon_\B(\PP)$. On the other hand, unrestricted minimization over $f$ would be uninformative, since $f=0$ never produces a loss. We therefore restrict attention to losing events that occur with probability at least $w$ for some preparation. This requirement is precisely $\lmax(P_f)\ge w$.

\begin{definition}[Random exclusion profile]
\label{def:random-exclusion-profile}
For $0<w\le1$, the random exclusion index of $\PP$ is
\begin{equation}
    E^w(\PP)
    :=\min_{f:\cX\to[0,1],\,\lmax(P_f)\ge w}\lmin(P_f)
    =\min_{F\in\Eff(\PP):\,\lmax(F)\ge w}\lmin(F).
    \label{eq:def-Ew}
\end{equation}
The indexed family $(E^w(\PP))_{0<w\le1}$ is called the random exclusion profile of $\PP$.
\end{definition}

The feasible set is nonempty and compact: it contains $f_x=w$ for every $x$, and is a closed subset of $[0,1]^{\cX}$. Hence the minimum is attained. Smaller values express a stronger exclusion capability at the probability threshold $w$.

\begin{proposition}[Basic properties of the random exclusion profile]
\label{prop:Ew-basic}
For every POVM $\PP=\{P_x\}_{x\in\cX}$, the following statements hold:
\begin{enumerate}
    \item $0\le E^w(\PP)\le w$ for every $0<w\le1$.
    \item $E^w(\PP)$ is nondecreasing in $w$.
    \item $\PP$ is trivial if and only if $E^w(\PP)=w$ for every $0<w\le1$.
    \item There exists $0<w\le1$ with $E^w(\PP)=0$ if and only if $\Eff(\PP)$ contains a nonzero singular effect, or equivalently, if and only if some nonzero $P_x$ is singular.
\end{enumerate}
\end{proposition}

\begin{proof}
Nonnegativity follows from positivity of the effects. For the upper bound, choose $f_x=w$, so that $P_f=w\1$. Monotonicity in $w$ follows because increasing $w$ restricts the feasible set.

If $\PP$ is trivial, then every $P_f$ is scalar, and the probability constraint forces $\lmin(P_f)=\lmax(P_f)\ge w$. The upper bound already proved gives $E^w(\PP)=w$. Conversely, if $\PP$ is nontrivial, some $P_{x'}$ is not proportional to the identity. Choosing its indicator function $f_x=\delta_{x',x}$ and $w=\lmax(P_{x'})>0$ gives
\[
    E^w(\PP)\le\lmin(P_{x'})<\lmax(P_{x'})=w.
\]

If $E^w(\PP)=0$, a minimizing effect $P_f$, which exists by compactness, is singular and satisfies $\lmax(P_f)\ge w>0$, so it is nonzero. Conversely, a nonzero singular $P_f$ gives $E^w(\PP)=0$ at $w=\lmax(P_f)$.

It remains to relate singular effects in $\Eff(\PP)$ to the individual effects $P_x$. Suppose $P_f$ is nonzero and singular, and choose a nonzero vector $v\in\ker P_f$. Then
\[
    0=\langle v,P_fv\rangle
     =\sum_x f_x\langle v,P_xv\rangle.
\]
Since every summand is nonnegative, positivity implies $P_xv=0$ whenever $f_x>0$. Moreover, since $P_f\ne0$, at least one such $P_x$ is nonzero, and hence singular. The converse follows because each $P_x$ belongs to $\Eff(\PP)$.
\end{proof}

Thus perfect exclusion is possible at some positive $w$ if and only if at least one nonzero effect is rank deficient. Rank deficiency of an effect is also the condition under which label assignments of an unlabeled POVM can be conclusively excluded~\cite{Ragini_Ziman_2024,Vijayan_Ragini_2026}. Null outcome labels do not affect the profile, because they do not affect $\Eff(\PP)$.

If $\Tr{\sigma P_f}=0$, occurrence of the event associated with $P_f$ excludes $\sigma$ conclusively. More generally, $\Tr{\sigma P_f}$ is the probability of incorrectly rejecting $\sigma$, while $\Tr{\rho P_f}\ge w$ requires the event to occur with probability at least $w$ for an alternative preparation $\rho$.

\subsection{Hypothesis-testing identity}

For states $\rho,\sigma$ and $0\le\varepsilon<1$, define
\begin{equation}
    \Delta_H^\varepsilon(\rho\Vert\sigma)
    :=\min_{0\le F\le\1:\,\Tr{\rho F}\ge1-\varepsilon}\Tr{\sigma F}.
    \label{eq:delta-h}
\end{equation}
This is the smallest type-II error among tests whose type-I error is at most $\varepsilon$. The corresponding hypothesis-testing relative entropy~\cite{Buscemi_Datta_2010,Wang_Renner_2012} is
$D_H^\varepsilon(\rho\Vert\sigma):=-\log\Delta_H^\varepsilon(\rho\Vert\sigma)$, with $-\log0=+\infty$.

\begin{theorem}[Random exclusion as optimized hypothesis testing]
\label{thm:hypothesis-testing}
For every POVM $\PP$ and $0<w\le1$,
\begin{equation}
    E^w(\PP)
    =\min_{\rho,\sigma\in\state(\hi)}
      \Delta_H^{1-w}\!\left(\M_\PP(\rho)\middle\Vert\M_\PP(\sigma)\right).
    \label{eq:Ew-hypothesis-testing}
\end{equation}
Equivalently,
\begin{equation}
    -\log E^w(\PP)
    =\max_{\rho,\sigma\in\state(\hi)}
      D_H^{1-w}\!\left(\M_\PP(\rho)\middle\Vert\M_\PP(\sigma)\right).
    \label{eq:Ew-hypothesis-testing-entropy}
\end{equation}
\end{theorem}

\begin{proof}
Both output states are diagonal in the outcome basis $\{\ket{x}\}_{x\in\cX}$, so replacing any effect $F$ in \cref{eq:delta-h} by its diagonal part preserves the relevant probabilities. We may therefore take $F=\sum_x f_x\ketbra{x}{x}$ with $0\le f_x\le1$, which gives
\[
    \Tr{\M_\PP(\rho)F}=\Tr{\rho P_f},
    \qquad
    \Tr{\M_\PP(\sigma)F}=\Tr{\sigma P_f}.
\]
The right-hand side of \cref{eq:Ew-hypothesis-testing} is consequently
\[
    \min_{\rho,\sigma,f:\,\Tr{\rho P_f}\ge w}\Tr{\sigma P_f}.
\]
For a fixed $f$, a feasible $\rho$ exists exactly when $\lmax(P_f)\ge w$, while minimization over $\sigma$ gives $\lmin(P_f)$. The last expression thus equals $E^w(\PP)$. Applying $-\log$ gives \cref{eq:Ew-hypothesis-testing-entropy}.
\end{proof}

\subsection{Monotonicity of the statistical hierarchy}

A classical post-processing cannot enlarge the effect range: if $\Q$ is a classical post-processing of $\PP$, then every effect $Q_g$ coincides with some $P_f$, obtained by absorbing the post-processing into the classical effect, so that $\Eff(\Q)\subseteq\Eff(\PP)$.
Unlike the set of indicator functions underlying $\epsilon_\B$, the set of all classical effects is closed under this absorption.
Together with the spectral contraction under positive unital maps, this gives monotonicity under both operations.

\begin{theorem}[Monotonicity of the random exclusion profile]
\label{thm:Ew-monotone}
Let $0<w\le1$.
\begin{enumerate}
    \item If $\Phi$ is positive and unital and $\PP'=\{\Phi(P_x)\}_{x\in\cX}$, then
    \[
        E^w(\PP')\ge E^w(\PP).
    \]
    \item If $\Q$ is obtained from $\PP$ by an arbitrary classical post-processing,
    \[
        Q_y=\sum_x\mu(y|x)P_x,
    \]
    then
    \[
        E^w(\Q)\ge E^w(\PP).
    \]
\end{enumerate}
\end{theorem}

\begin{proof}
For the first statement, take any $f$ feasible for $E^w(\PP')$. Since $P'_f=\Phi(P_f)$, positivity and unitality imply
\[
    \lmax(P_f)\ge\lmax(P'_f)\ge w,
    \qquad
    \lmin(P'_f)\ge\lmin(P_f)\ge E^w(\PP).
\]
Minimizing the left-hand side over the feasible $f$ proves the claim.

For the second statement, every classical effect $g:\cY\to[0,1]$ satisfies
\[
    Q_g\equiv\sum_y g_yQ_y
       =\sum_x\left(\sum_y\mu(y|x)g_y\right)P_x
       =P_f,
    \qquad
    f_x:=\sum_y\mu(y|x)g_y.
\]
Since $\mu(\cdot|x)$ is a probability distribution, $0\le f_x\le1$. Hence
\[
    \Eff(\Q)\subseteq\Eff(\PP),
\]
and the minimization for $\Q$ is over a subset of the effects available for $\PP$.
\end{proof}

The second statement also follows from the data-processing inequality for hypothesis testing through \cref{eq:Ew-hypothesis-testing}.

\subsection{The exclusion preorder and binary statistical ranges}

The random exclusion profile defines a comparison between measurements that need not have the same outcome alphabet or act on the same Hilbert space.

\begin{definition}[Exclusion preorder]
\label{def:exclusion-preorder}
For POVMs $\PP$ and $\Q$, write $\Q\preceq_{\mathrm{excl}}\PP$ if
\begin{equation}
    E^w(\Q)\ge E^w(\PP)
    \qquad\text{for every }0<w\le1.
    \label{eq:exclusion-preorder}
\end{equation}
\end{definition}

\begin{corollary}
\label{cor:excl_preorder}
If $\Q$ is obtained from $\PP$ by quantum pre-processing and classical post-processing, then $\Q\preceq_{\mathrm{excl}}\PP$.
\end{corollary}

To understand what this necessary condition retains, consider the binary POVM
\[
    f\circ\PP:=\{P_f,\1-P_f\}.
\]
Its range is a line segment whose endpoints are determined by the extremal eigenvalues:
\begin{equation}
    \Rng(f\circ\PP)
    =\left\{(t,1-t):\lmin(P_f)\le t\le\lmax(P_f)\right\}.
    \label{eq:binary_range}
\end{equation}
Thus range inclusion for two binary post-processings is exactly a comparison of the smallest and largest probabilities of their designated events.

\begin{proposition}[Binary range characterization]
\label{prop:exclusion-binary-ranges}
For POVMs $\PP=\{P_x\}_{x\in\cX}$ and $\Q=\{Q_y\}_{y\in\cY}$, the following statements are equivalent:
\begin{enumerate}
    \item $\Q\preceq_{\mathrm{excl}}\PP$.
    \item For every classical effect $g:\cY\to[0,1]$, there exists a classical effect $f:\cX\to[0,1]$ such that
    \[
        \Rng(g\circ\Q)\subseteq\Rng(f\circ\PP).
    \]
\end{enumerate}
\end{proposition}

\begin{proof}
Suppose first that $\Q\preceq_{\mathrm{excl}}\PP$, and fix $g$. If $Q_g=0$, the choice $f=0$ gives the required inclusion. Otherwise, set $w:=\lmax(Q_g)>0$ and choose $f$ attaining $E^w(\PP)$. Then
\[
    \lmax(P_f)\ge w=\lmax(Q_g),
\]
while
\[
    \lmin(P_f)=E^w(\PP)
    \le E^w(\Q)
    \le\lmin(Q_g).
\]
By \cref{eq:binary_range}, these two inequalities give the desired range inclusion.

Conversely, fix $0<w\le1$ and choose $g$ attaining $E^w(\Q)$. By assumption, there exists $f$ such that
\[
    \lmin(P_f)\le\lmin(Q_g),
    \qquad
    \lmax(P_f)\ge\lmax(Q_g)\ge w.
\]
Hence $f$ is feasible for $E^w(\PP)$, and
\begin{equation}
    E^w(\PP)\le\lmin(P_f)\le\lmin(Q_g)=E^w(\Q).
    \label{eq:binary-exclusion-comparison}
\end{equation}
Since $w$ was arbitrary, $\Q\preceq_{\mathrm{excl}}\PP$ follows.
\end{proof}

The matching post-processing is allowed to depend on the binary test, so this comparison does not yet require a single simulation of all outcomes of $\Q$. The question is therefore whether a refinement to more than two outcomes can enforce this consistency.

\section{Simulation hierarchies of POVMs}
\label{sec:Simulation hierarchy of POVMs}

We extend the binary comparison to more outcomes, first at the level of statistical ranges, obtaining positive unital maps on the operator systems spanned by the effects. We then include an ancillary measurement and obtain a characterization of simulation by quantum pre-processing and classical post-processing. The distinction between these two constructions concerns complete positivity; as we shall see, it disappears at the binary level or, more generally, when the target POVM is commutative.

The hierarchies constructed here follow the pattern of the statistical comparison of experiments, where experiments are ordered by their performance in decision problems with a prescribed number of decisions. Testing problems are the binary case, which suffices for pairs of probability distributions~\cite{Blackwell_1953,Torgersen_1970} but not, in general, for pairs of quantum states~\cite{Jencova_2012}. In the quantum versions of the theory, two channels with a common input are compared through the statistics that tests on their outputs produce, for every input, with the number of test outcomes grading the comparison~\cite{Buscemi_2016,Buscemi_2018}. Comparison at all cardinalities characterizes the existence of a quantum statistical morphism~\cite{Buscemi_2012}, namely a linear map that is well behaved on the range of the first channel but need not extend to a positive map on the whole operator algebra~\cite{Matsumoto_2014}. When the target channel has classical output, or when the comparison includes an ancillary system, the corresponding criteria yield a post-processing channel~\cite{Buscemi_2012,Buscemi_2016,Buscemi_2018}. For convertibility under general convex families of channel transformations see Ref.~\cite{jencova-comparison-2021}.

Throughout this section, let $\PP=\{P_x\}_{x\in\cX}$ and $\Q=\{Q_y\}_{y\in\cY}$ act on $\hi_\PP$ and $\hi_\Q$, respectively. These Hilbert spaces need not have the same dimension. We denote by $\operatorname{Post}_k(\PP)$ the set of all $k$-outcome classical post-processings of $\PP$:
\begin{equation}
\begin{split}
    \operatorname{Post}_k(\PP)
    &:=\left\{\nu\circ\PP:\nu(a|x)\geq0,
    \ \sum_{a=1}^k\nu(a|x)=1\right\},\\
    \nu\circ\PP&:=\{P_a^\nu\}_{a=1}^k,
    \qquad P_a^\nu:=\sum_x\nu(a|x)P_x.
\end{split}
\label{eq:pos-postprocessing-set}
\end{equation}
Null effects are allowed, so that a POVM with fewer than $k$ outcomes can be regarded as a $k$-outcome POVM by adjoining zero effects.

\subsection{Comparison by statistical ranges}
\label{subsec:pos-simulation-hierarchy}

\begin{definition}[$k$-outcome positive simulation preorder]
\label{def:pos-k-simulation}
For $k\geq2$, we write $\Q\preceq_{\mathrm{pos}}^{(k)}\PP$ if, for every $\mathsf R\in\operatorname{Post}_k(\Q)$, there exists $\mathsf S\in\operatorname{Post}_k(\PP)$ such that
\begin{equation}
    \mathscr R(\mathsf R)\subseteq\mathscr R(\mathsf S).
    \label{eq:pos-k-range-inclusion}
\end{equation}
\end{definition}

The order of the quantifiers is essential: once the post-processing of $\Q$ is chosen, a single post-processing of $\PP$ must reproduce its entire statistical range. That post-processing cannot depend on the input state. Reflexivity and transitivity follow directly from this definition, since classical post-processings compose and set inclusion is transitive. Hence $\preceq_{\mathrm{pos}}^{(k)}$ is indeed a preorder.

For $k=2$, a stochastic matrix is specified by a classical effect $f_x=\nu(1|x)$, with $\nu(2|x)=1-f_x$. Its post-processing is therefore the binary POVM $\{P_f,\1-P_f\}$. By the binary-range characterization of the exclusion preorder established in the preceding section,
\begin{equation}
    \Q\preceq_{\mathrm{pos}}^{(2)}\PP
    \quad\Longleftrightarrow\quad
    \Q\preceq_{\mathrm{excl}}\PP.
    \label{eq:pos-binary-exclusion}
\end{equation}

To understand the higher levels, we use the following characterization of range inclusion, which is Proposition~7.1 of Ref.~\cite{buscemi-2005-clean-POVMs}. We include a proof because the domain of the resulting map is relevant to the subsequent discussion.

\begin{lemma}[Range inclusion and positive unital maps]
\label{lem:pos-range-map}
Let $\mathsf S=\{S_a\}_{a=1}^k$ and $\mathsf R=\{R_a\}_{a=1}^k$ be POVMs on possibly different Hilbert spaces. The following statements are equivalent:
\begin{enumerate}[label=\textup{(\roman*)}]
    \item $\mathscr R(\mathsf R)\subseteq\mathscr R(\mathsf S)$.
    \item There exists a positive unital linear map
    \[
        \Phi:\operatorname{span}_{\mathbb C}\{S_a\}_a
        \longrightarrow\operatorname{span}_{\mathbb C}\{R_a\}_a
    \]
    such that $\Phi(S_a)=R_a$ for every $a$.
\end{enumerate}
If such a map exists, it is unique on its stated domain.
\end{lemma}

\begin{proof}
Suppose first that (i) holds. For every state $\tau$ on $\hi_\mathsf R$, choose a state $\rho_\tau$ on $\hi_\mathsf S$ satisfying
\begin{equation}
    \Tr{\tau R_a}=\Tr{\rho_\tau S_a}
    \qquad\text{for every }a.
    \label{eq:pos-matching-states}
\end{equation}
If $\sum_a e_aS_a=0$ for some $e\in \mathbb{C}^k$, then $\Tr{\tau\sum_a e_aR_a}=0$ for every $\tau$, and hence $\sum_a e_aR_a=0$. Consequently,
\[
    \Phi\left(\sum_a c_aS_a\right):=\sum_a c_aR_a, \qquad c\in\mathbb{C}^k,
\]
defines a linear map. It is unital because both POVMs sum to the identity. If $A\geq0$ belongs to its domain, Eq.~\eqref{eq:pos-matching-states} gives
\[
    \Tr{\tau\Phi(A)}=\Tr{\rho_\tau A}\geq0
\]
for every $\tau$, so $\Phi(A)\geq0$. Its uniqueness follows because the effects $S_a$ span the domain.

Conversely, suppose that (ii) holds. For each real vector $c=(c_a)_{a=1}^k$, positivity and unitality imply
\[
    \lambda_{\max}\left(\sum_a c_aR_a\right)
    =\lambda_{\max}\left(\Phi\left(\sum_a c_aS_a\right)\right)
    \leq\lambda_{\max}\left(\sum_a c_aS_a\right).
\]
The two sides are the support functions of $\mathscr R(\mathsf R)$ and $\mathscr R(\mathsf S)$, respectively, since
\[
    \max_{p\in\mathscr R(\mathsf R)}\sum_a c_ap_a
    =\lambda_{\max}\left(\sum_a c_aR_a\right),
\]
and similarly for $\mathsf S$. Both ranges are compact and convex. The support-function inequality therefore implies $\mathscr R(\mathsf R)\subseteq\mathscr R(\mathsf S)$ by the separating-hyperplane theorem.
\end{proof}

\begin{theorem}[Nesting and positive-map characterization]
\label{thm:pos-hierarchy}
The positive simulation preorders have the following properties:
\begin{enumerate}[label=\textup{(\roman*)}]
    \item For every $k\geq2$,
    \[
        \Q\preceq_{\mathrm{pos}}^{(k+1)}\PP
        \quad\Longrightarrow\quad
        \Q\preceq_{\mathrm{pos}}^{(k)}\PP.
    \]
    \item Set $m:=\max\{2,|\cY|\}$. The following conditions are equivalent:
    \begin{enumerate}[label=\textup{(\alph*)}]
        \item $\Q\preceq_{\mathrm{pos}}^{(m)}\PP$;
        \item $\Q\preceq_{\mathrm{pos}}^{(k)}\PP$ for every $k\geq2$;
        \item There exist a stochastic matrix $\nu(y|x)$ and a positive unital map
        \[
            \Phi:\operatorname{span}_{\mathbb C}\{S_y\}_y
            \longrightarrow\operatorname{span}_{\mathbb C}\{Q_y\}_y,
            \qquad S_y:=\sum_x\nu(y|x)P_x,
        \]
        satisfying $\Phi(S_y)=Q_y$ for every $y$.
    \end{enumerate}
\end{enumerate}
\end{theorem}

\begin{proof}
To prove (i), take $\mathsf R\in\operatorname{Post}_k(\Q)$ and adjoin a null effect, obtaining
\[
    \widetilde{\mathsf R}=(R_1,\ldots,R_k,0)\in\operatorname{Post}_{k+1}(\Q).
\]
The assumption provides $\widetilde{\mathsf S}\in\operatorname{Post}_{k+1}(\PP)$ whose range contains that of $\widetilde{\mathsf R}$. Merge the last two output labels of both POVMs. This leaves $\mathsf R$ on the left and gives a POVM $\mathsf S\in\operatorname{Post}_k(\PP)$ on the right; applying the same stochastic map to the two ranges preserves their inclusion. Thus $\mathscr R(\mathsf R)\subseteq\mathscr R(\mathsf S)$.

For (ii), condition (b) immediately implies (a). If $|\cY|\geq2$, condition (a) applied to the identity post-processing of $\Q$ gives $\mathsf S=\nu\circ\PP$ with $\mathscr R(\Q)\subseteq\mathscr R(\mathsf S)$. Lemma~\ref{lem:pos-range-map} then gives (c). In the single-outcome case, $\Q=\{\1\}$, and (c) holds directly by taking $S_1=\1$ and $\Phi(c\1)=c\1$.

Finally, assume (c) and fix an arbitrary stochastic matrix $\mu$ with $k$ output labels. Set
\[
    R_a:=\sum_y\mu(a|y)Q_y,\qquad
    T_a:=\sum_y\mu(a|y)S_y.
\]
Then $\mathsf T\in\operatorname{Post}_k(\PP)$ and $\Phi(T_a)=R_a$. The restriction of $\Phi$ to $\operatorname{span}_{\mathbb C}\{T_a\}_a$ is positive and unital, so Lemma~\ref{lem:pos-range-map} yields $\mathscr R(\mathsf R)\subseteq\mathscr R(\mathsf T)$. Since $\mu$ and $k$ were arbitrary, (b) follows.
\end{proof}

Once $k$ is large enough that no outcome of the target POVM $\Q$ needs to be merged, i.e., $k\ge|\cY|$, the comparison is characterized by classical post-processing followed by a positive unital map on the operator system. To interpret it as quantum pre-processing, however, one must still establish complete positivity and extension to the full operator algebra.

\begin{remark}[Commutative target measurements]
\label{rem:pos-commutative-target}
If the effects of $\Q$ commute, the map in Theorem~\ref{thm:pos-hierarchy}(ii)(c) is completely positive. Indeed, simultaneous diagonalization gives
\[
    \Phi(A)=\sum_j\varphi_j(A)\Pi_j, \qquad 
    \varphi_j(A):=\frac{\operatorname{Tr}[\Pi_j\Phi(A)]}{\operatorname{Tr}\Pi_j},
\]
where the $\Pi_j$ are the joint spectral projections of $\Q$ and each $\varphi_j$ is a positive unital scalar-valued functional on $\operatorname{span}_{\mathbb C}\{S_y\}_y$. Such functionals are completely positive, hence so is $\Phi$. Arveson's extension theorem then extends $\Phi$ to a completely positive unital map on $\mathcal L(\hi_\PP)$~\cite{Arveson_1969}. This is Theorem~7.3 in Ref.~\cite{buscemi-2005-clean-POVMs}, applied to the pair $(\mathsf{S},\Q)$ with $\mathsf{S}=\nu\circ \PP$, and it parallels the fact that, in the comparison of channels, statistical morphisms can be replaced by post-processing channels when the target has commuting outputs~\cite{Buscemi_2016,Buscemi_2018}.

The same conclusion does not follow merely from the commutativity of the source effects $\mathsf{S}_y$, as is the case, for instance, when $\PP$ is commutative. In that case their span is an operator subsystem of a commutative algebra, but it need not be the whole algebra, and a positive map on that subsystem need not admit a positive extension. The distinction, and a counterexample, are discussed in Ref.~\cite{buscemi-2005-clean-POVMs}; an analogous phenomenon for families of states is exhibited in Ref.~\cite{Matsumoto_2014}. Thus commutativity of the source and commutativity of the target play different roles in this comparison.
\end{remark}

\subsection{Comparison with an ancillary measurement}
\label{subsec:cp-simulation-hierarchy}

Complete positivity can be tested by retaining correlations with an ancillary system. For a comparison whose target is $\Q$, fix an ancillary Hilbert space $\mathcal K\simeq\hi_\Q$ and an informationally complete POVM $\mathsf M=\{M_z\}_{z\in\mathcal Z}$ on $\mathcal K$. Here informational completeness means that the complex span of its effects is $\mathcal L(\mathcal K)$. The same ancillary POVM is used on both sides of every range inclusion in the following definition.

\begin{definition}[$k$-outcome completely positive simulation preorder]
\label{def:cp-k-simulation}
For $k\geq2$, we write $\Q\preceq_{\mathrm{CP}}^{(k)}\PP$ if, for every $\mathsf R\in\operatorname{Post}_k(\Q)$, there exists $\mathsf S\in\operatorname{Post}_k(\PP)$ such that
\begin{equation}
    \mathscr R(\mathsf R\otimes\mathsf M)
    \subseteq\mathscr R(\mathsf S\otimes\mathsf M).
    \label{eq:cp-k-range-inclusion}
\end{equation}
The range on the left is evaluated over states on $\hi_\Q\otimes\mathcal K$, whereas the range on the right is evaluated over states on $\hi_\PP\otimes\mathcal K$.
\end{definition}

The next lemma shows that the definition is independent of the chosen informationally complete POVM. It is the range criterion of Theorem~7.2 in Ref.~\cite{buscemi-2005-clean-POVMs}, stated here for possibly different input dimensions. We give a direct proof that also constructs the pre-processing channel.

\begin{lemma}[Ancillary range inclusion and completely positive maps]
\label{lem:cp-range-map}
Let $\mathsf S=\{S_a\}_{a=1}^k$ and $\mathsf R=\{R_a\}_{a=1}^k$ act on $\hi_\mathsf S$ and $\hi_\mathsf R$, respectively. Fix $\mathcal K\simeq\hi_\mathsf R$ and an informationally complete POVM $\mathsf M=\{M_z\}_{z\in\mathcal Z}$ on $\mathcal K$. The following statements are equivalent:
\begin{enumerate}[label=\textup{(\roman*)}]
    \item $\mathscr R(\mathsf R\otimes\mathsf M)\subseteq\mathscr R(\mathsf S\otimes\mathsf M)$.
    \item There exists a completely positive unital map
    \[
        \Phi:\operatorname{span}_{\mathbb C}\{S_a\}_a
        \longrightarrow\operatorname{span}_{\mathbb C}\{R_a\}_a
    \]
    satisfying $\Phi(S_a)=R_a$ for every $a$.
    \item There exists a completely positive unital map
    \[
        \widehat\Phi:\mathcal L(\hi_\mathsf S)
        \longrightarrow\mathcal L(\hi_\mathsf R)
    \]
    satisfying $\widehat\Phi(S_a)=R_a$ for every $a$.
\end{enumerate}
The map in (ii) is unique on its stated domain; its extension in (iii) need not be unique. Whenever these conditions hold, range inclusion also holds with any ancillary POVM on any finite-dimensional ancillary system.
\end{lemma}

\begin{proof}
The proof essentially follows the argument presented in~\cite{Buscemi_2016}; nonetheless, we include it here for completeness.

We first prove (i)$\Rightarrow$(iii). Write $d=\dim\hi_\mathsf R$ and fix matching orthonormal bases of $\hi_\mathsf R$ and $\mathcal K$. Let
\[
    \ket{\Omega}=\frac{1}{\sqrt d}\sum_{j=1}^d\ket{j}\otimes\ket{j}.
\]
By (i), there is a state $\sigma$ on $\hi_\mathsf S\otimes\mathcal K$ such that
\begin{equation}
    \Tr{\sigma(S_a\otimes M_z)}
    =\bra{\Omega}(R_a\otimes M_z)\ket{\Omega}
    =\frac1d\Tr{R_a^{\mathsf T}M_z}
    \label{eq:cp-maxent-matching}
\end{equation}
for every $a,z$. Informational completeness therefore implies
\begin{equation}
    \operatorname{Tr}_{\hi_\mathsf S}
    \bigl[(S_a\otimes\1)\sigma\bigr]
    =\frac1d R_a^{\mathsf T}.
    \label{eq:cp-choi-marginals}
\end{equation}
Summing over $a$ gives $\operatorname{Tr}_{\hi_\mathsf S}\sigma=\1_{\mathcal K}/d$. Hence $d\sigma$ is the Choi operator~\cite{Choi_1975} of the completely positive trace-preserving map
\[
    \Lambda(B)
    :=d\operatorname{Tr}_{\mathcal K}
    \bigl[(\1\otimes B^{\mathsf T})\sigma\bigr],
    \qquad B\in\mathcal L(\hi_\mathsf R).
\]
Equation~\eqref{eq:cp-choi-marginals} gives $\Lambda^\dagger(S_a)=R_a$. Taking $\widehat\Phi=\Lambda^\dagger$ proves (iii).

Restriction gives (iii)$\Rightarrow$(ii), while Arveson's extension theorem gives (ii)$\Rightarrow$(iii)~\cite{Arveson_1969}; unitality is preserved because the identity belongs to the domain. To prove (iii)$\Rightarrow$(i), let $\Lambda=\widehat\Phi^\dagger$. For every bipartite state $\tau$,
\[
    \Tr{\tau(R_a\otimes M_z)}
    =\Tr{(\Lambda\otimes\mathrm{id})(\tau)(S_a\otimes M_z)}.
\]
Since $\Lambda\otimes\mathrm{id}$ maps states to states, this proves range inclusion. The same argument applies to any ancillary system and any ancillary POVM. Finally, uniqueness on the operator system follows because the $S_a$ span that system.
\end{proof}

Consequently, $\Q\preceq_{\mathrm{CP}}^{(k)}\PP$ is equivalent to the following statement: for every $k$-outcome post-processing $\mathsf R$ of $\Q$, there exist a $k$-outcome post-processing $\mathsf S$ of $\PP$ and a completely positive unital map on the full operator algebra satisfying $\Phi(S_a)=R_a$ for every $a$. This description proves reflexivity and transitivity, since such maps and stochastic post-processings compose. It also shows that
\begin{equation}
    \Q\preceq_{\mathrm{CP}}^{(k)}\PP
    \quad\Longrightarrow\quad
    \Q\preceq_{\mathrm{pos}}^{(k)}\PP.
    \label{eq:cp-implies-pos}
\end{equation}

We now consider simulation by one quantum pre-processing channel and one classical post-processing matrix. More precisely, a channel $\Lambda:\mathcal L(\hi_\Q)\to\mathcal L(\hi_\PP)$ and a stochastic matrix $\nu(y|x)$ simulate $\Q$ from $\PP$ if
\begin{equation}
    Q_y=\Lambda^\dagger\left(\sum_x\nu(y|x)P_x\right)
    \qquad\text{for every }y.
    \label{eq:cp-physical-simulation}
\end{equation}
In the Schr\"odinger picture, this means applying $\Lambda$ to the input state, measuring $\PP$, and then applying $\nu$ to the observed classical outcome. 
Note that a mixed strategy of pairs $(\Lambda, \nu)$ by shared randomness need not yield a simulation of the form \cref{eq:cp-physical-simulation}.
The criterion below concerns simulation with a fixed pair $(\Lambda,\nu)$ as in \cref{eq:cp-physical-simulation}.

\begin{theorem}[Nesting and complete characterization of simulation]
\label{thm:cp-hierarchy}
The completely positive simulation preorders have the following properties:
\begin{enumerate}[label=\textup{(\roman*)}]
    \item For every $k\geq2$,
    \[
        \Q\preceq_{\mathrm{CP}}^{(k+1)}\PP
        \quad\Longrightarrow\quad
        \Q\preceq_{\mathrm{CP}}^{(k)}\PP.
    \]
    \item Set $m:=\max\{2,|\cY|\}$. The following conditions are equivalent:
    \begin{enumerate}[label=\textup{(\alph*)}]
        \item $\Q\preceq_{\mathrm{CP}}^{(m)}\PP$;
        \item $\Q\preceq_{\mathrm{CP}}^{(k)}\PP$ for every $k\geq2$;
        \item There exist a channel $\Lambda$ and a stochastic matrix $\nu$ satisfying Eq.~\eqref{eq:cp-physical-simulation}.
    \end{enumerate}
\end{enumerate}
\end{theorem}

\begin{proof}

For (i), as in the proof of \cref{thm:pos-hierarchy}, we adjoin a null effect to $\R\in\operatorname{Post}_k(\Q)$ to obtain $\widetilde\R\in\operatorname{Post}_{k+1}(\Q)$, take $\widetilde{\mathsf S}\in\operatorname{Post}_{k+1}(\PP)$ provided by the assumption.
We then merge the last two output labels of both $\widetilde\R$ and $\widetilde{\mathsf S}$, leaving the output labels of the ancillary POVM unchanged.
Since the merging acts as the same classical post-processing on both sides, the inclusion $\Rng(\widetilde\R\otimes\mathsf M)\subseteq\Rng(\widetilde{\mathsf S}\otimes\mathsf M)$ provided by the assumption yields $\Rng(\R\otimes\mathsf M)\subseteq\Rng(\mathsf S\otimes\mathsf M)$, where $\mathsf S\in\operatorname{Post}_k(\PP)$ denotes the merged POVM.

For (ii), (b)$\Rightarrow$(a) is immediate. If $|\cY|\geq2$, apply (a) to the identity post-processing of $\Q$, obtaining a post-processing $\mathsf S=\nu\circ\PP$ with
\[
    \mathscr R(\Q\otimes\mathsf M)
    \subseteq\mathscr R(\mathsf S\otimes\mathsf M).
\]
Lemma~\ref{lem:cp-range-map} supplies a completely positive unital map $\widehat\Phi:\mathcal L(\hi_\PP)\to\mathcal L(\hi_\Q)$ satisfying $\widehat\Phi(S_y)=Q_y$. Its adjoint is a channel, giving (c). If $|\cY|=1$, the target is the trivial POVM $\{\1\}$, which is simulated by sending every output of $\PP$ to its sole label, with any pre-processing channel.

Conversely, suppose (c) holds and choose any stochastic matrix $\mu$ with $k$ output labels. Define
\[
    \chi(a|x):=\sum_y\mu(a|y)\nu(y|x),
\]
then it holds that
\[
    (\mu\circ\Q)_a
    =\Lambda^\dagger\bigl((\chi\circ\PP)_a\bigr).
\]
Lemma~\ref{lem:cp-range-map} therefore gives the required ancillary range inclusion for every $\mu$ and every $k$, proving (b).
\end{proof}

When the comparison uses fewer outcomes than the target POVM, i.e., $k<|\cY|$, the pre-processing channel $\Lambda$ and the classical post-processing $\nu$ simulating $\R$ may depend on the chosen $\R\in\operatorname{Post}_k(\Q)$; once all outcomes of $\Q$ are retained, i.e., $k\ge|\cY|$, a single pair $(\Lambda,\nu)$ must simulate $\Q$ itself, and hence all of its post-processings.

\subsection{The binary case}
\label{subsec:cp-binary-level}

Although complete positivity is an additional requirement for general operator systems, a binary measurement has a simpler structure due to its inherent commutativity. Its statistical range is an interval, and the endpoints of that interval suffice to construct a measure-and-prepare simulation. The following lemma is essentially Theorem~7.5 of Ref.~\cite{buscemi-2005-clean-POVMs}, slightly extended to allow for different Hilbert spaces and to provide an explicit measure-and-prepare, and hence entanglement-breaking~\cite{Horodecki_2003}, realization.

\begin{lemma}[Binary range inclusion admits a channel simulation]
\label{lem:cp-binary-simulation}
Let $\mathsf S=\{A,\1-A\}$ and $\mathsf R=\{B,\1-B\}$ be binary POVMs on possibly different Hilbert spaces. If $\mathscr R(\mathsf R)\subseteq\mathscr R(\mathsf S)$, then there exists a completely positive unital map
\[
    \Phi:\mathcal L(\hi_\mathsf S)\longrightarrow\mathcal L(\hi_\mathsf R)
\]
satisfying $\Phi(A)=B$. Its adjoint can be chosen to be a measure-and-prepare channel.
\end{lemma}

\begin{proof}
Write $\alpha=\lambda_{\min}(A)$ and $\beta=\lambda_{\max}(A)$. Range inclusion implies $\alpha\1\leq B\leq\beta\1$. If $\alpha<\beta$, choose states $\rho_-$ and $\rho_+$ supported on the eigenspaces of $A$ corresponding to $\alpha$ and $\beta$, respectively, and define the effect
\[
    F:=\frac{B-\alpha\1}{\beta-\alpha}.
\]
Then
\begin{equation}
    \Phi(C):=\Tr{\rho_-C}(\1-F)+\Tr{\rho_+C}F
    \label{eq:cp-binary-measure-prepare}
\end{equation}
is completely positive and unital on the full operator algebra, and
\[
    \Phi(A)=\alpha(\1-F)+\beta F=B.
\]
Its adjoint measures $\{\1-F,F\}$ and prepares $\rho_-$ or $\rho_+$ according to the outcome. If $\alpha=\beta$, then $A=\alpha\1$ and $B=\alpha\1$, and the map $\Phi(C)=\Tr{\rho C}\1$, with any state $\rho$, has the required properties.
\end{proof}

\begin{corollary}[Coincidence of the binary comparisons]
\label{cor:cp-pos-binary}
For all POVMs $\PP$ and $\Q$,
\begin{equation}
    \Q\preceq_{\mathrm{CP}}^{(2)}\PP
    \quad\Longleftrightarrow\quad
    \Q\preceq_{\mathrm{pos}}^{(2)}\PP
    \quad\Longleftrightarrow\quad
    \Q\preceq_{\mathrm{excl}}\PP.
    \label{eq:cp-pos-excl-binary}
\end{equation}
In particular, if $\Q$ has two outcomes, its random exclusion profile gives a complete criterion for simulation from $\PP$ by quantum pre-processing and classical post-processing.
\end{corollary}

\begin{proof}
The implication from the completely positive comparison to the positive comparison follows from Eq.~\eqref{eq:cp-implies-pos}. Conversely, apply Lemma~\ref{lem:cp-binary-simulation} to every binary range inclusion appearing in the positive comparison, and then apply Lemma~\ref{lem:cp-range-map}. The equivalence with the exclusion preorder is Eq.~\eqref{eq:pos-binary-exclusion}. The last statement follows from Theorem~\ref{thm:cp-hierarchy}, since the binary level retains all outcomes of a binary target.
\end{proof}

\section{Discussion}\label{sec:Discussion}

The algebraic exclusion profile bounds classical communication cost, while the random exclusion profile compares the binary statistical capabilities of measurements. The distinction between binary comparison and simulation of the whole measurement concerns the consistency of the simulating operations: an ordering of random exclusion profiles guarantees a separate simulation of every binary post-processing of the target, whereas retaining all target outcomes, together with the ancillary range criterion, requires a single simulation of the whole POVM.

Several questions remain. Relating approximate range inclusion and smoothed exclusion profiles to the minimum simulation error would give quantitative criteria for approximate simulation. Such criteria would be particularly relevant near measurements with rank-deficient effects, since an event with small positive probability may go unobserved in a finite sample. Allowing the quantum pre-processing and the classical post-processing to coordinate their choices through shared randomness would permit convex mixtures of the transformations considered here. An exclusion-based characterization of the resulting simulation preorder would connect the present hierarchy with convex theories of channel conversion~\cite{jencova-comparison-2021}. For a given nonbinary target POVM, one may also ask how many outcomes the completely positive comparison must retain to characterize simulation with a fixed pair, and in particular when the binary level suffices. Finally, the communication interpretation raises the question of when the bounds derived from the algebraic exclusion profile determine the minimum number of classical messages needed to reproduce the full conditional statistics of the forbidden-output task with shared randomness.

\section*{Acknowledgments}
K.T. acknowledges support from JST SPRING Grant No.~JPMJSP2125. J.K. acknowledges support from MEXT Quantum Leap Flagship Program (MEXT QLEAP) Grant No.~JPMXS0120319794. F.B. acknowledges support from MEXT Quantum Leap Flagship Program (MEXT QLEAP) Grant No.~JPMXS0120319794 and from JSPS KAKENHI Grants No.~23K03230 and~26K00621.

\section*{Author contribution statement}
K.T. led the research, with supporting contributions from J.K. F.B. supervised the project. 
In addition, ChatGPT (OpenAI, Astra 6 and Sol 6.1) and Claude (Anthropic, Opus 5.5) were used, under the author’s direction, to assist with literature searches, the examination of mathematical arguments, and the preparation and revision of the manuscript. The authors take full responsibility for the content of this work.

\bibliographystyle{alpha}
\newcommand{\etalchar}[1]{$^{#1}$}

\end{document}